\pdfoutput=1	

\documentclass[10pt,letterpaper]{article}
\usepackage{opex3}

\begin{document}



\newcommand{\change}[2]{#1}
\newcommand{\addition}[1]{#1}

\title{Imaging with parallel ray-rotation sheets}

\author{Alasdair C.\ Hamilton and Johannes Courtial}


\address{Department of Physics and Astronomy, Faculty of Physical Sciences, University of Glasgow, Glasgow G12~8QQ, United~Kingdom}

\email{a.c.hamilton@physics.gla.ac.uk} 



\begin{abstract}
A ray-rotation sheet consists of miniaturized optical components
that function -- ray optically -- as a homogeneous medium that rotates the local direction of transmitted light rays around the sheet normal by an arbitrary angle [A.\ C.\ Hamilton \textit{et al.}, arXiv:0809.2646 (2008)].
Here we show that two or more parallel ray-rotation sheets perform imaging between two planes.
The image is unscaled and un-rotated.
No other planes are imaged.
When seen through parallel ray-rotation sheets, planes that are not imaged appear rotated\change{.}{was ``, whereby the rotation angle changes with the ratio between the observer's and the object plane's distance from the sheets.''}
\end{abstract}

\ocis{(110.0110) Imaging systems;
(110.2990) Image formation theory;
(160.1245) Artificially engineered materials;
(240.3990) Micro-optical devices
}

\bibliography{/Users/johannes/Documents/work/library/Johannes}
\bibliographystyle{osajnl}

\section{Introduction\label{introduction-section}}

Sheets composed of miniaturized optical components can perform \change{interesting}{was ``surprising''} ray-optical transformations.
Such \change{sheets can}{was ``systems may''} act very differently from \change{}{was ``that of''} the \change{optical component on its own}{was ``original, isolated optical component''}.
\change{A sheet comprising a pair of confocal}{was ``Confocal''} lenslet arrays, for example, can act approximately like the interface between optical media with different refractive indices, including negative refractive indices \cite{Courtial-2008a}.
Another example is a Dove-prism sheet, an array of Dove prisms that flips one transverse component of the local direction of transmitted light rays (for example the $x$ component) \cite{Hamilton-Courtial-2008a}.
If two parallel Dove-prism sheets, one immediately behind the other, are arranged such that their flip directions are orthogonal (``crossed Dove-prism sheets''), they flip both transverse components of the local light-ray direction -- they act like the interface between optical media with opposite refractive indices \cite{Courtial-Nelson-2008}.
If the two sheets' flip axes are not orthogonal, the two Dove-prism sheets rotate the direction of transmitted light rays by twice the angle between the flip axes \cite{Hamilton-et-al-2009}.
Such ray-rotation sheets are without wave-optical analog \cite{Hamilton-Courtial-2009}.

When seen through a ray-rotation sheet, light rays originating from a given point appear to come from a different direction \cite{Hamilton-et-al-2009}.
The direction change is such that a point light source is not geometrically imaged, unless the point light source lies in the plane of the ray-rotation sheet, when the sheet merely changes the direction in which light rays leave the source.
\change{In this sense, a}{was ``A''} ray-rotation sheet \change{}{was ``therefore''} performs trivial imaging of the sheet plane into itself.

We investigate here geometric imaging with two or more parallel \addition{idealized} ray-rotation sheets, separated by finite distances.
We show that such parallel \addition{idealized} ray-rotation sheets perform \change{unusual}{used to be ``non-trivial, and very unusual,''} imaging:
two planes are imaged into each other with magnification $+1$ (and no image rotation); no other plane is imaged.
\addition{An optical system consisting of two or more parallel ray-rotation sheets has no optical axis.
Therefore, the imaging quality does not vary across the image plane.
An increase in aperture size in principle leads to no loss of ray-optical imaging quality.
These imaging characteristics are quite different from those that can be achieved using (conventional or Fresnel) lenses.}


\section{\label{ray-rotation-sheets-section}Ray-rotation sheets}

\begin{figure}
\begin{center}
\includegraphics{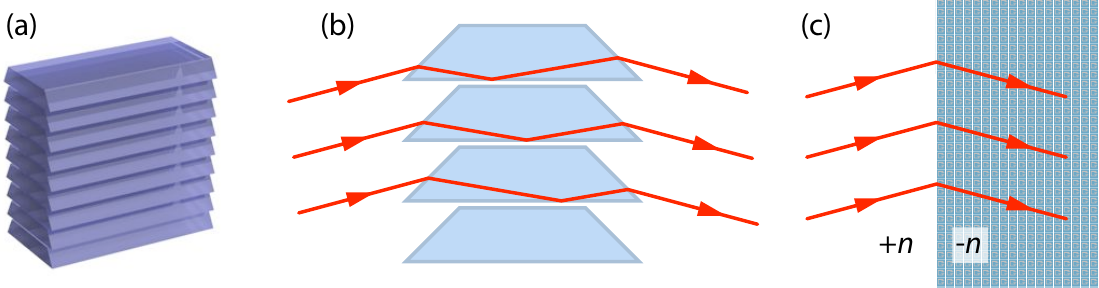}
\end{center}
\caption{\label{ray-flipping-fig}Optics of a single Dove-prism array.  (a)~A Dove-prism array is formed by a stack of Dove prisms that form a sheet.
(b)~Each individual Dove prism flips one of the transverse direction components of light rays passing through it, here the vertical component.
(c)~If the Dove prisms are miniaturized, the overall effect of a Dove-prism sheet on the flipped direction of transmitted light rays is equivalent to that of a planar interface between two materials with opposite refractive indices, $+n$ and $-n$.}
\end{figure}


In all of this paper apart from this section we treat ray-rotation sheets in terms of their \addition{idealized} effect on transmitted light rays, namely rotation around the local sheet normal.
Here we \change{review}{was ``discuss''} briefly one particular design of a ray-rotation sheet -- two  almost coplanar Dove-prism sheets \change{}{removed ``(Fig.\ \ref{ray-flipping-fig}(a))''} -- and \change{how transmission through such sheets can be described to a very good approximation as pure light-ray rotation}{was ``how it causes this rotation''} \addition{\cite{Hamilton-et-al-2009}}.

A stack of very thin Dove prisms can form a Dove-prism sheet, as shown in Fig.\ \ref{ray-flipping-fig}(a).
Each individual Dove prism flips one transverse component of the direction of transmitted light rays (Fig.\ \ref{ray-flipping-fig}(b)).
In general, the transverse positions at which a light ray enters and exits the prism are different -- the prism offsets the light rays, whereby the size of the offset is of the order of the size of the prism aperture.
By miniaturizing the Dove prisms, this offset can be made almost arbitrarily small while still maintaining the Dove-prism sheet's direction-flipping property.
The effect is then equivalent to that of the interface with two materials with opposite refractive indices, but only as far as the flipped ray-direction component is concerned.
(The effect of two Dove-prism sheets in parallel planes but with crossed flip directions is a flipping of both transverse ray-direction components, which in turn is ray-optically equivalent to an opposite-refractive-index interface.)

\begin{figure}
\begin{center}
\includegraphics{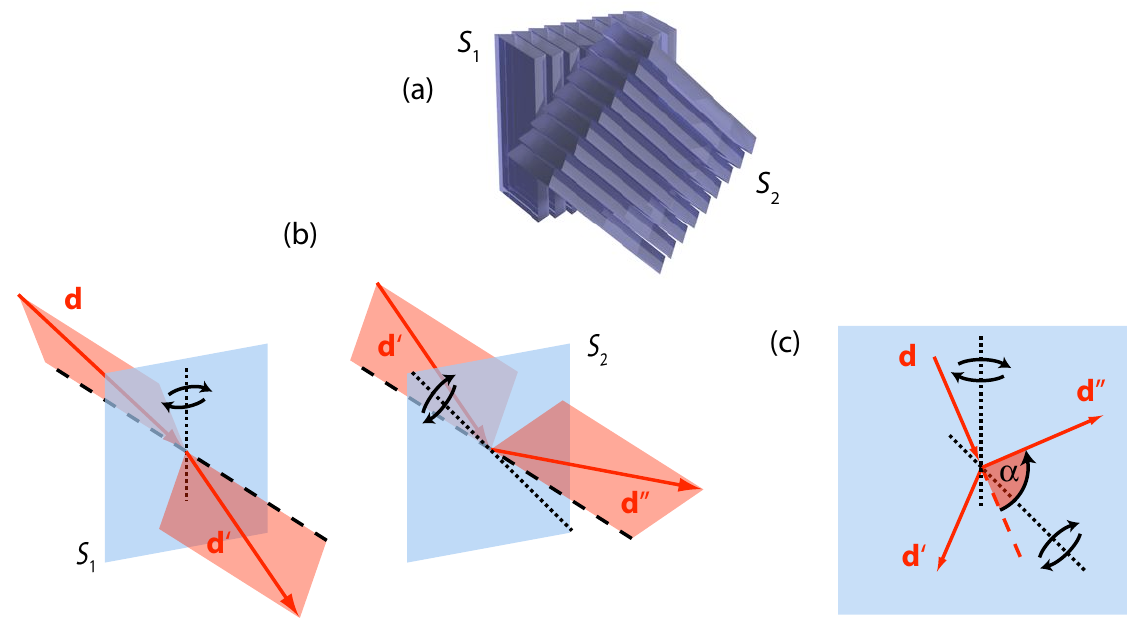}
\end{center}
\caption{\label{ray-rotation-fig}
\change{Example of local light-ray rotation with Dove-prism sheets.
(a)~Structure of two parallel Dove-prism sheets, $S_1$ and $S_2$, that are rotated with respect to each other by $\alpha/2 = 45^\circ$.
The two sheets individually flip the transverse direction of transmitted light rays, but with respect to different axes.
(b)~Successive flipping of a light-ray direction, $\mathbf{d}$, first by sheet $S_1$ (left frame), resulting in the intermediate direction $\mathbf{d}^\prime$, then by sheet $S_2$, resulting in direction $\mathbf{d}^{\prime \prime}$.
Dove-prism sheets are represented by semi-transparent, light-blue, squares (seen from a 3D position from where they appear as parallelograms).
Light-ray directions are represented by red arrows; the planes containing the light-ray directions and the local Dove-prism-sheet normal (dashed black lines) are indicated by semi-transparent red rectangles.
The flip of the transverse light-ray direction is with respect to an axis in the sheet plane (dotted line).
(c)~A plot of the transverse light-ray directions (orthographic projection of the light-ray directions into the sheet plane or any other transverse plane) reveals that the two successive flips of the transverse ray direction are equivalent to rotation through an angle $\alpha$.}{changed quite a bit}}
\end{figure}

A pair of Dove-prism sheets, one immediately in front of the other, can form a ray-rotation sheet.
The Dove-prism sheets are in parallel planes, but the directions of the Dove prisms in the two sheets -- and with them the two sheets' flip axes -- are rotated with respect to each other around a sheet normal (Fig.\ \ref{ray-rotation-fig}(a)).
\change{By analogy with}{was ``In analogy to the''} image rotation\change{, which}{was ``that''} can be achieved by flipping the image twice with respect to different axes, ray rotation can be achieved by flipping the ray direction twice \change{(Fig.\ \ref{ray-rotation-fig}(b)).}{was ``. This is shown in Fig.\ \ref{ray-rotation-fig}(b).''}
In both cases the rotation angle is twice the angle between the flip axes.
In order to achieve a ray-rotation angle $\alpha$, the two Dove-prism sheets therefore have to be rotated with respect to each other by $\alpha/2$.

A ray-rotation sheet based on Dove prisms suffers from a number of imperfections.
These include a limited field of view; the small, but nevertheless non-zero, ray offset mentioned above; and diffraction effects, particularly in the case of small Dove prisms.
A few of these issues are studied in more detail elsewhere~\cite{Hamilton-et-al-2009}.

\addition{In the remainder of this paper we consider idealized ray-rotation sheets.}



\section{\label{twoRayRotationSheet-sec}Geometric imaging with ray-rotation sheets}

If an optical system re-directs all the light rays from a point light source at point $L$ such that all the light rays that have passed through the optical system intersect again in another point, $L^\prime$, then the optical system images $L$ into $L^\prime$.
$L$ is called the object, $L^\prime$ its image.
Both $L$ and $L^\prime$ can be real or virtual: in the former case the actual light rays intersect, in the latter case their continuations.

\begin{figure}
\begin{center}
\includegraphics{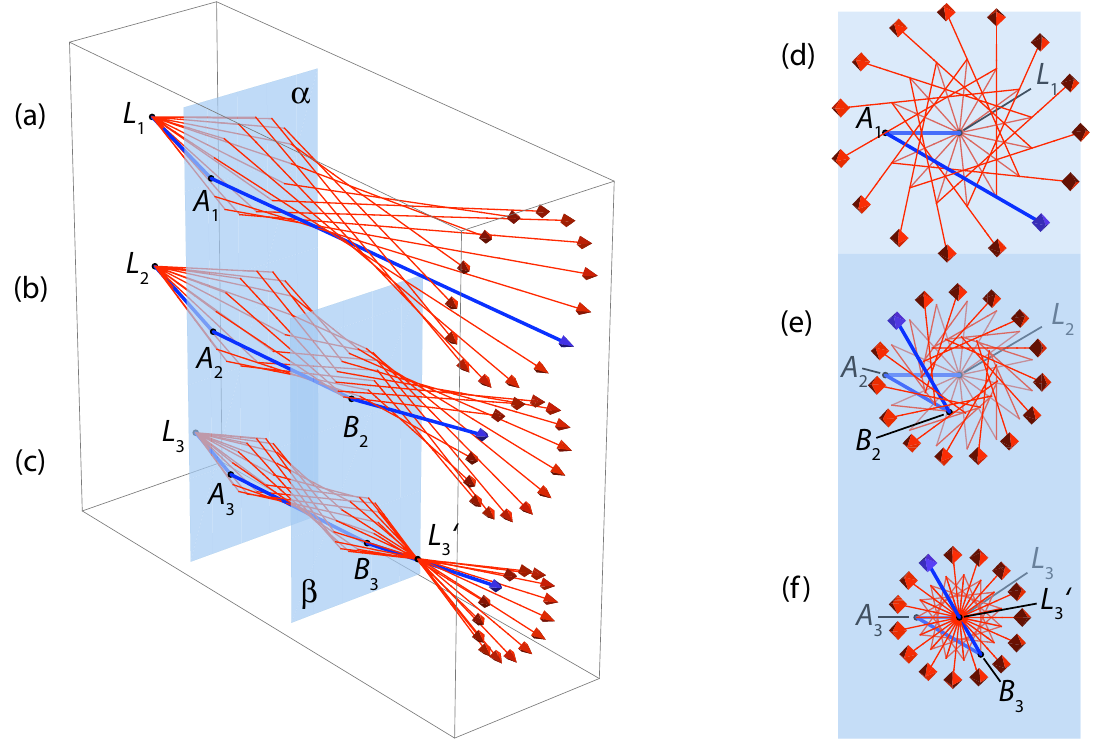}
\end{center}
\caption{\label{side-views-figure}
Trajectories of light rays that originate from three point light sources, $L_1$ to $L_3$, and pass through ray-rotation sheets.
All light-ray trajectories are shown in red, apart from the trajectory of one light ray from each light source, which is highlighted in blue.
The ray-rotation sheets are shown in light blue; the first rotates the direction of transmitted light rays through an angle $\alpha$ around the sheet normal, the second through an angle $\beta$.
In the case of a single ray-rotation sheet, light rays from a point light source not located in the sheet plane generally do not intersect again after passage through the sheet~(a).
The same is true for passage through two ray-rotation sheets~(b), unless the light source is located in the one transverse plane that is imaged by the two sheets, in which case all the light rays originating from the light source, $L_3$, intersect again in a point $L_3^\prime$~(c).
The points at which the highlighted trajectories intersect the first sheet are marked $A_1$ to $A_3$, those where they intersect the second sheet are $B_2$ and $B_3$.
Frames (d), (e) and (f) show the orthographic projections into a transverse plane of the light-ray trajectories respectively shown in (a), (b) and (c).
The figure is drawn for $\alpha = \beta = 150^\circ$.}
\end{figure}

Fig.\ \ref{side-views-figure}(a) shows a cone of light rays leaving a point light source, $L_1$, and passing through a ray-rotation sheet with rotation angle $\alpha$.
After passage through the sheet the rays form a twisted bundle; no two rays in the bundle intersect, so the sheet does not image $L$.
This is in fact typical of \addition{individual} ray-rotation sheets.
Except in special cases (ray-rotation angles $\alpha = 0^\circ$ and $\alpha = 180^\circ$), \change{a ray-rotation sheet does}{was ``ray-rotation sheets do''} not image point light sources in any plane other than \change{}{was ``-- trivially --''} the sheet plane, which is imaged again into the sheet plane\addition{, in the sense discussed in section \ref{introduction-section}}.
\change{}{removed ``(That imaging the sheet plane into itself is indeed trivial is illustrated by the fact that any thin, but randomly patterned, sheet of glass performs imaging of the sheet plane into the sheet plane.
An example of such a sheet is a toilet window.)'' to save a bit of space}

Fig.\ \ref{side-views-figure}(b) shows the trajectories of a cone of light rays from another point light source, $L_2$, passing through two parallel ray-rotation sheets with rotation angles $\alpha$ and $\beta$, respectively.
Like the typical single-ray-rotation-sheet case, passage through two ray-rotation sheets results in a twisted bundle of non-intersecting rays.
The sheets do not produce an image of the light source, and this situation is again typical.

Fig.\ \ref{side-views-figure}(c) shows the special case in which imaging takes place.
This time the light source, $L_3$, is positioned at a distance from the first ray-rotation sheet such that, after passage through both sheets, all light rays originating from $L_3$ intersect again in the same point, $L_3^\prime$, the image of $L_3$.

Fig.\ \ref{side-views-figure}(d), (e) and (f) show these three cases from a different perspective.
These plots are orthographic projections of the ray bundles and sheets into a transverse plane.
From the projections of the twisted ray bundles in the first two, non-imaging, cases, Fig.\ \ref{side-views-figure}(d) and (e), it is immediately clear that there is no point at which all the rays from the same point light source intersect.
The third, imaging, case (Fig.\ \ref{side-views-figure}(f)) is different.
When following the projection of the highlighted ray trajectory it can be seen that, after transmission through both ray-rotation sheets, it passes through the projection of the point light source.
As this is the case not only for the highlighted light ray, but for light rays leaving the point light source in any direction, \emph{all} light rays intersect at the point where the projections of the ray and the point light source intersect.

\begin{figure}
\begin{center}
\includegraphics{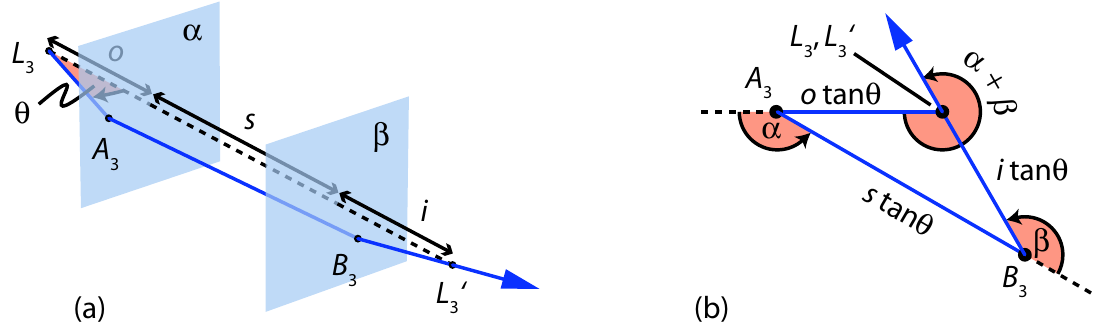}
\end{center}
\caption{ \label{imaging-trajectory-figure}Geometry of a light-ray trajectory in the imaging case.
The trajectory \addition{(solid dark-blue arrow)} is the same as that highlighted in Fig.\ \ref{side-views-figure}(c) and (f).
\change{(a)~Three-dimensional representation.
The ray-rotation sheets (rotation angles $\alpha$ and $\beta$, respectively) are shown in light blue.
$\theta$ is the angle between the ray trajectory and the normal to both ray-rotation sheets.
(The angle is shown in the form of a red segment.)
The dashed line is the sheet normal through the point light source, $L_3$, and its image, $L_3^\prime$.
The distances $o$ (between $L_3$ and the first sheet), $s$ (between the two sheets), and $i$ (between the second sheet and $L_3^\prime$) are marked by double-sided arrows.
(b)~Orthographic projection into a transverse plane.
The angles $\alpha$ and $\beta$ by which the direction of the projected ray changes are simply the angles by which the two ray-rotation sheets rotate the ray around the local sheet normal.
The relevant angles are marked as red segments.}{was ``(a)~Three-dimensional representation;
(b)~orthographic projection into a transverse plane.''}}
\end{figure}

We now study this quantitively.
We concentrate on one light-ray trajectory in the imaging case (Fig.\ \ref{imaging-trajectory-figure}(a)).
First we notice that, as a ray-rotation sheet rotates the light-ray direction around the sheet normal, it does not change the angle between the light ray and the sheet normal.
If two or more successive ray-rotation sheets share the same sheet normal, the angle between the ray trajectory and the direction of the normal remains constant.
This is the case in our parallel ray-rotation sheets.
We call the angle with the sheet normal $\theta$.
The transverse and longitudinal distances, respectively $\Delta r$ and $\Delta z$, travelled by a light ray that is inclined by an angle $\theta$ with respect to the sheet normal (which we choose to be the $z$ direction), are then related through the equation
\begin{equation}
\frac{\Delta r}{\Delta z} = \tan \theta.
\label{polar-angle-equation}
\end{equation}

For the light-ray trajectory's orthographic projection into a transverse plane (Fig.\ \ref{imaging-trajectory-figure}(b)), equation (\ref{polar-angle-equation}) implies that the side lengths of the triangle $L_3 A_3 B_3$ (where $A_3$ and $B_3$ are the points where the trajectory intersects the first and second sheet, respectively -- see Figs \ref{side-views-figure} and \ref{imaging-trajectory-figure}) are the product of the corresponding $z$ distance and $\tan \theta$.
We call the distance between the point light source, $L_3$, and the first sheet the object distance, $o$, we call the separation between the first and second sheet $s$,
and we call the distance between the second sheet and the image of $L_3$, $L_3^\prime$, the image distance, $i$.
Basic trigonometry then leads to the following object distance, given two sheets with ray-rotation angles $\alpha$ and $\beta$, separated by $s$:
\begin{equation}
\label{o-equation}
o = - \frac{\sin\beta}{\sin(\alpha+\beta)} s.
\end{equation}
Similarly, the image distance is
\begin{equation}
\label{i-equation}
i = - \frac{\sin\alpha}{\sin(\alpha+\beta)} s.
\end{equation}

\begin{figure}
\begin{center}
\includegraphics{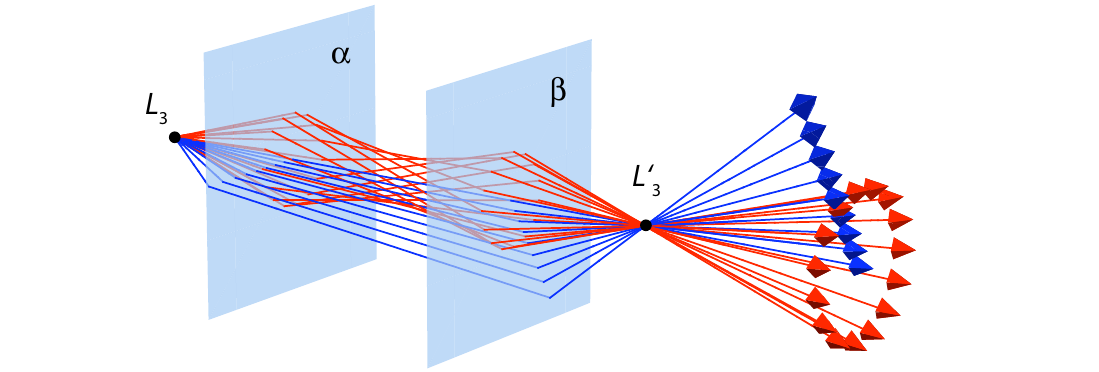}
\end{center}
\caption{ \label{imaging-trajectories-cone-and-fan-figure}\addition{Focussing of light rays with different angles $\theta$ with the common sheet normal by a pair of parallel ray-rotation sheets (shown as light blue, semi-transparent, squares).
The red light rays leave the point light source, $L_3$, with an angle $\theta = 17^\circ$ with respect to the sheet normal; they are the light rays shown in Fig.\ \ref{side-views-figure}(c), which is calculated for the same ray-rotation angles ($\alpha = \beta = 150^\circ$), the same separation between the ray-rotation sheets, and the same position of $L_3$.
The blue light rays leave $L_3$ at angles $\theta = 0^\circ, 5^\circ, 10^\circ, ... 40^\circ$ with respect to the sheet normal.
All light rays intersect again in the same point, $L_3^\prime$, the image of $L_3$.
To facilitate appreciation of the 3D structure of the ray trajectories, the supplementary material contains a movie (MPEG-4, 1.4MB) of the same arrangement seen from different viewing positions.}}
\end{figure}

There are a number of things to note about these equations for object and image distance.
Firstly, they are independent of the angle $\theta$.
This means that rays that leave the light source at any angle $\theta$ with the common sheet normal are imaged at the point $L_3^\prime$ (provided, of course, they pass through both ray-rotation sheets) \addition{ (Fig.\ \ref{imaging-trajectories-cone-and-fan-figure})}.
So $L_3^\prime$ is the geometrical image of $L_3$.
\addition{The independence from the angle $\theta$, together with the fact that no approximations were used in finding that $L_3$ is imaged into $L_3^\prime$, implies that idealized ray-rotation sheets perform ray-optically perfect imaging.
Specifically, an increase in aperture size leads to no loss of ray-optical imaging quality.
Obviously, combinations of particular implementations of ray-rotation sheets are limited by how closely the particular implementations approximate idealized ray-rotation sheets.}

\change{Secondly}{was ``Thirdly''}, equations (\ref{o-equation}) and (\ref{i-equation}) hold not only for the particular position of light source $L_3$, but for any light-source position in the plane a distance $o$ in front of the first \addition{ray-rotation} sheet\addition{, which is then imaged to a corresponding position in the plane a distance $i$ behind the second ray-rotation sheet}.
This means the entire plane is imaged.
\addition{(Similarly, the fact that any point light source that is a distance other than $o$ in front of the first ray-rotation sheet results in a twisted bundle of non-intersecting rays means that no other plane is imaged.)}
As the orthographic projections into a transverse plane of any light source and its image coincide, the image of the plane is the same size as the object and upright, so the magnification of the imaging process is $M = +1$.

\begin{figure}[ht]
\begin{center}
\includegraphics{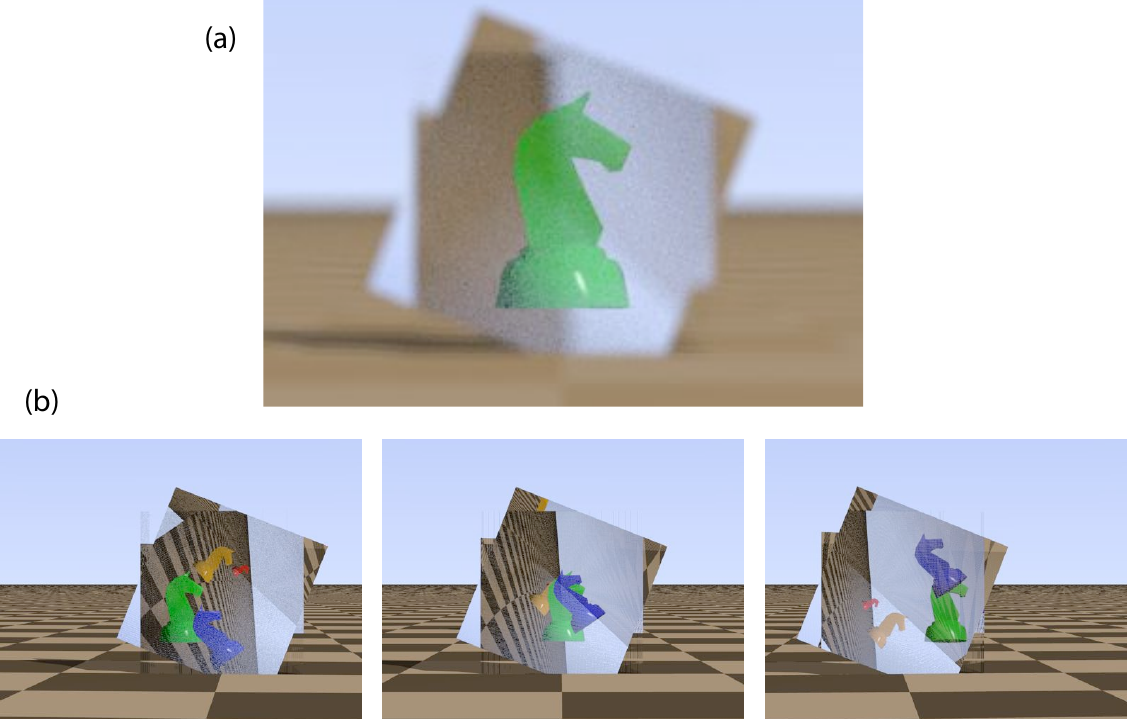}
\end{center}
\caption{\label{simulation-figure}\addition{Simulated} view through a pair of ray-rotation sheets.
The ray-rotation angle of each sheet is $137^\circ$; in units of the floor-tile length, the sheets are separated by a distance $s=5$, so the object and image distance are $o=i=3.42$.
(a)~\change{A green chess piece is placed behind the ray-rotation sheets, in the object plane.
The rendering parameters are chosen such that only the image plane is in focus -- the picture is a simulation of a photo taken with a camera that is focussed on the image plane and that has a finite-size aperture.
As expected, a sharp image of the chess piece can be seen in the image plane.}
{was ``Green chess piece in the object plane.
The rendering parameters are chosen such that only the image plane is in focus.''}
(b)~\change{Additional}{was ``Several''} (upright) chess pieces \addition{are placed} in different planes behind the ray-rotation sheets.
\change{As before, the}{was ``The''} green chess piece is in the object plane (at distance $o$), the other pieces are not (\addition{they are} at distances $o/2$ (blue), $2 o$ (orange), and $4 o$ (red)).
This time the rendering parameters were chosen so that all planes are rendered in focus\addition{; the frames are now simulations of photos taken with a camera with a negligibly small aperture}.
\change{Therefore all chess pieces can be seen in sharp focus, but note}{was ``(Note''} that there is still only one object and one image plane that are being imaged into each other by the pair of ray-rotation sheets.\change{}{was ``)''}
\change{To demonstrate this,}{was ``The''} three frames were calculated for different camera positions, namely (from left to right) right of the central sheet normal, on the central sheet normal, and left of the central sheet normal.
\addition{This horizontal camera movement makes the chess pieces that are not in the object plane appear to move vertically, which cannot be reconciled with these chess pieces being imaged, as discussed in the main text.}
The figure was calculated using the ray-tracing software POV-Ray \cite{POV-Ray}, which simulated transmission through the detailed structure of two ray-rotation sheets, each consisting of two Dove-prism sheets \cite{Hamilton-et-al-2009}, each in turn consisting of 1000 Dove prisms. 
}
\end{figure}

Figure \ref{simulation-figure}(a) shows a simulation of a chess piece in the object plane of two ray-rotation sheets, seen through the sheets.
\addition{Note that all frames in Fig.\ \ref{simulation-figure} are ray-tracing simulations of the view through the detailed structure of ray-rotation sheets consisting of Dove-prism sheets (section \ref{ray-rotation-sheets-section}).}
The (simulated) camera is focussed on the image plane, and a sharp image of the chess piece can be seen there, as expected.

Fig.\ \ref{simulation-figure}(b) shows a simulation of chess pieces in different planes, \addition{again} seen through two ray-rotation sheets.
This time the camera is not focussed on any particular plane, so all pieces appear sharp.
That only the chess piece in the object plane is actually imaged can be seen when looking at the same scene from \change{different angles}{was ``a different angle''}:  if the camera is moved sideways, the pieces that are not imaged appear to move sideways with respect to each other -- as images at different distances should --, but they also appear to move up and down, as no stationary object or image would do when the camera is moved sideways.
\addition{(Note that the camera movement simulated in Fig.\ \ref{simulation-figure}(b) can be interpreted as seeing the same scene from different positions within a much larger aperture.
The fact that any objects that are not imaged appear to move would then lead to those objects appearing blurred on a photo taken with a lens with such a much larger aperture.)}

\change{Thirdly, the absence of any center of magnification (as the magnification is $M = +1$) or any rotation axis (contrary to what might be expected from transmission through a combination of ray-rotation sheets, the image is not rotated with respect to the object) implies that there is no optical axis.
This also follows from the lack of any preferred axis in each of the constituent ray-rotation sheets.
}{was ``Unlike what might be expected from transmission through ray-rotation sheets, the image is not rotated with respect to the object.''}

\addition{These first three characteristics of imaging with parallel ray-rotation sheets are quite different from those of any system of lenses, irrespective of whether or not the lenses are thick or thin and whether or not the lenses are in the form of Fresnel lenses, which bear a resemblance to Dove-prism sheets.
Specifically, any system of lenses 
images not only points in one plane, but points in any position (unless the lens system contains a planar diffuser, in which case only the planes into which the diffuser plane is imaged are imaged into each other, whereby the diffuser plane contains an intermediate image);
and any lens has an optical axis, which leads to a preferred axis.
The lack of an optical axis makes ray-rotation-sheet imaging more akin to imaging with a ``perfect'' lens formed by the planar interface between media with refractive indices of equal magnitude but opposite sign \cite{Pendry-2000};
in fact, transmission through just such an interface is the same as idealized light-ray rotation through $180^\circ$ \cite{Hamilton-et-al-2009}.
In other ways, for example imaging of only a single plane, imaging with parallel ray-rotation sheets differs from imaging with a perfect lens.}


\change{Fourthly}{was ``Secondly''}, equations (\ref{o-equation}) and (\ref{i-equation}) can give positive and negative object and/or image distances, respectively corresponding to real and virtual objects/images.

Finally, as the magnification between object and image plane is $+1$, these planes are principal planes \cite{Smith-1978}.
The effect of the two sheets is then that of a single sheet with ray-rotation angle $\alpha + \beta$ (Fig.\ \ref{imaging-trajectory-figure}(b)) and principal planes in the object and image planes.
In this way, we therefore recover the result that a single ray-rotation sheet trivially images the sheet plane into itself.
This argument can also easily be extended to $N$ ray-rotation sheets with rotation angles $\alpha_1, \alpha_2, ..., \alpha_N$ and separations $s_1, s_2, ..., s_{N-1}$: they act like a single ray-rotation sheet with a ray-rotation angle $\alpha = \sum_j \alpha_j$ and principal planes (that is, object and image planes with magnification $M=+1$) at distances
\begin{equation}
o= - \frac{\sum_j s_j \sin \left( \sum_{k=j+1}^N \alpha_k \right)}{\sin \alpha}, \quad
i = - \frac{\sum_j s_j \sin \left( \sum_{k=1}^j \alpha_k \right)}{\sin \alpha}
\end{equation}
in front of the first sheet and behind the last sheet, respectively.

\section{\label{conclusions-section}Conclusions}

\change{Combinations of ray-rotation sheets can perform geometrical imaging that is, in principle, perfect and which is different from the geometrical imaging performed by simple combinations of lenses.}{We have described here geometric imaging with parallel ray-rotation sheets.
Only two planes are imaged into each other, with magnification $M=+1$;
objects in any other plane appear rotated when seen through the sheets.}
\change{The unusual imaging properties of ray-rotation sheets}{was ``These properties are very unusual and''} might well find applications, for example in range finding.

\section*{Acknowledgments}
Many thanks to John Nelson for help with the POV-Ray simulations.
A.C.H.\ is supported by the UK's Engineering and Physical Sciences Research Council (EPSRC).
J.C.\ is a Royal Society University Research Fellow.

\end{document}